%% file: ms6.tex
\documentclass[12pt,preprint]{aastex}
\usepackage{color}

\def\rel{{\rm rel}}
\def\e{{\rm E}}
\def\au{{\rm AU}} 
\def\muas{{\mu\rm as}}
\def\kms{{\rm km}\,{\rm s}^{-1}}
\def\kpc{{\rm kpc}}

\def\rel{{\rm rel}}

\def\mas{{\rm mas}}
\def\masyr{{\rm mas\,yr^{-1}}}

\def\e{{\rm E}}

\begin{document}
\title{OGLE-2015-BLG-1459L: The Challenges of Exo-Moon Microlensing}
\input author

\begin{abstract} 

We show that dense OGLE and KMTNet $I$-band survey data require
four bodies (sources plus lenses) to explain the
microlensing light curve of OGLE-2015-BLG-1459.  However, these can
equally well consist of three lenses
and one source (3L1S), two lenses and two sources (2L2S) or one lens
and three sources (1L3S).  In the 3L1S and 2L2S interpretations, the host is
a brown dwarf and the dominant companion is a Neptune-class planet,
with the third body (in the 3L1S case) being a Mars-class
object that could have been a moon of the planet.  In the 1L3S 
solution, the light curve anomalies are explained by 
a tight (five stellar radii) low-luminosity binary source that is 
offset from the principal source of the event by $\sim 0.17\,\au$.
These degeneracies are resolved in favor of the 1L3S solution by
color effects derived from comparison to MOA data, which 
are taken in a slightly different ($R/I$) passband. 
{To enable current and future ($WFIRST$) surveys to routinely
  characterize exomoons and distinguish among such exotic systems
  requires an observing strategy that includes both a cadence faster
  than 9 min$^{-1}$ and observations in a second band on a similar
  timescale.}

\end{abstract}


\section{{Introduction}
\label{sec:intro}}

The eight planets of the Solar System harbor an amazing diversity
of moons.  Two planets have no moons, while Jupiter has four major
moons that Johannes Kepler already realized constitute a mini-Solar System
obeying his Third Law.  Some moons are mostly ice while others
are entirely composed of rock.  A few have atmospheres, lakes, geysers,
and other features.  Some were captured and others formed in situ.
It is even speculated that some moons harbor life.

While the number of confirmed or highly-probable exo-planets is
several hundred times greater than the eight Solar System planets,
the situation for exo-moons is the reverse: no clear 
discoveries (but see \citealt{teachey17}).  
Exo-moons generate almost zero signal in Doppler studies
of host stars because the barycenter of the planet-moon system follows
almost exactly the same orbit as an isolated planet.  The transit
method is more sensitive to exo-moons: Earth's moon would give rise
to a transit signal that is $\sim 7\%$ of Earth's.  However, since
the Earth signal is itself near the detection limit for a single transit
for the {\it Kepler} satellite, similar moons would be most easily detected
for very close-in planets that had many transits during the mission.
See \citet{kipping15} and references therein.

The microlensing method may have greater potential to detect
exo-moons.  Microlensing occurs when a massive body (star, planet, etc)
becomes closely aligned with a more distant source star.  The gravity of the
foreground object (lens) bends the source light and thereby magnifies it.
The changing magnification gives rise to a ``microlensing event''.
In the case of planet-star (or more complicated) systems, the light curve
can be complex, thus revealing the presence, geometry, and masses of
multiple components.

{The detectability of an isolated object (be it lunar, planetary,
  or stellar mass) is governed by microlensing's characteristic
angular scale, the Einstein radius}
\begin{equation}
\theta_\e = \sqrt{\kappa M \pi_\rel};
\qquad \kappa\equiv {4G\over c^2\au}\simeq 8.1{\mas\over M_\odot},
\label{eqn:thetaedef}
\end{equation}
where $M$ is the lens mass and
$\pi_\rel\equiv \au(D_L^{-1}-D_S^{-1})$ is the lens-source
relative parallax.  Hence, the cross section for microlensing 
detection scales $\theta_\e\sim M^{1/2}$.  

{If the lens consistes of two bodies, then in the limit that the separation between them is $s\gg 1$, the light curve
can appear as two isolated ``bumps''. In this case, the requirement for detecting the second bump (and second body) is that} the source
passes within roughly $\theta_{\e,p}=\sqrt{q}\theta_\e$ of the planet, 
where $q$ is the planet-star mass ratio, e.g., OGLE-2016-BLG-0263
\citep{ob160263}.  This mildly favors high-mass
planets but because there are more low-mass than high-mass planets, 
microlensing planet detections are almost uniform in $\log q$ \citep{ob160596}.
As the planet gets closer to the Einstein radius of the star, it gives rise
to a growing caustic structure (contours of formally infinite magnification
for point sources) that becomes much larger than its own
Einstein radius.  Hence, over half of detections have projected separations
very close to the star's Einstein radius, $0.8<s<1.25$ \citep{ob160596}. 
{The scaling of the caustics with $q$ is either $q^{1/3}$,
  $q^{1/2}$, or $q$ depending on the type of caustic \citep{gaudi12}.}

{An exomoon differes from a planet only in that it lies in the
  extreme low-mass regime.}
For a lens lying
halfway toward the Galactic bulge (where all microlensing planet
searches are conducted), $\theta_\e=1.7\,\muas(M/M_\oplus)^{1/2}$.
For exo-moons, this is likely to be smaller than the angular radius of
the source, typically $\theta_*\sim 0.5\,\muas$ {(e.g. $\theta_\e = 0.17\, \muas$ for $M = 0.01 M_{\oplus}$, $D_L = 4\,$ kpc, $D_S = 6\,$ kpc)}.  Thus, an isolated
exo-moon would magnify only a small fraction of the source, making it
difficult or impossible to detect.  However, just as the planet
generates a larger effect if it is close to the star's Einstein
radius, its moon can likewise have an outsized effect on the planet's
caustic.  
{Thus, one expects that exo-moons would be most easily detectable
  by distorting the caustic due to its host planet (although in
  principle they can also change the topology of the caustic, e.g. by
  adding extra cusps). The topology variations of three-body lens systems are discussed in detail by \citet{danek15a}, \citet{danek15b}, and {\citet{Song14}}.}
\citet{han02}, \citet{han08}, \citet{liebig10}, and \citet{chung16}
have studied exo-moon{-specific} features of microlensing light
curves and find a wide variety of light curve morphologies.

Here we investigate the microlens OGLE-2015-BLG-1459L and show that
based on the $I$-band light curve alone,
it is an exo-moon candidate.  However, we also show that there
are two alternate solutions.  In the first, there is only a host
and planet but no exo-moon.  The additional anomaly that could
be attributed to an exo-moon is then attributed to a second source
(companion to the primary source).  In the second, both the
main and secondary anomalies are attributed to the companion sources,
so that there is a triple source system but only a single lens.
We then resolve these degeneracies in favor of the triple-source solution
by measuring the evolving color
of the event.  Finally, we discuss the wider implications of this
event for the practical study of exo-moons with microlensing.

\section{{Observations}
\label{sec:obs}}

\begin{figure}
\centering
\includegraphics[width=16cm]{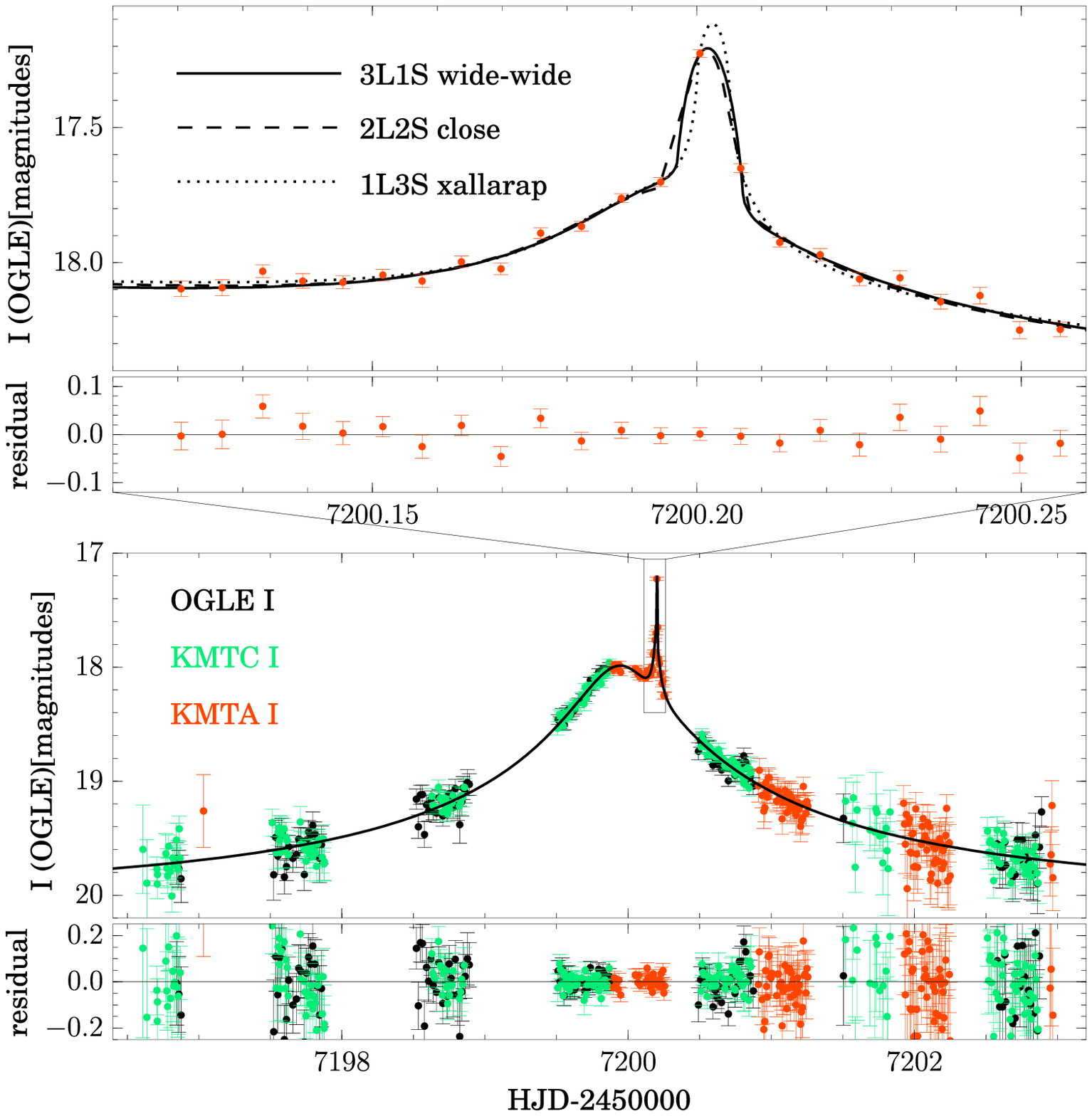}
\caption{\noindent OGLE-2015-BLG-1459 light curve together
with three models that fit the $I$-band data from OGLE and KMTNet 
equally well.   One model (3L1S) contains three bodies: a
brown dwarf host, a Neptune-class planet and a Mars-class 
object that may orbit the ``Neptune''.  The second model (2L2S)
contains a brown dwarf host and a Neptune-class planet.  The third
model (1L3S) has a single lens that microlenses a triple-source system.
Actually, there are
four variants of the planet/moon model and
two variants of the planet model, but the remaining solutions look
almost identical to those shown (wide-wide and close, respectively).
}
\label{fig:lc}
\end{figure}

On 2015 June 25 (HJD$^\prime=$HJD-2450000 =
7199.1) the Optical Gravitational Lensing Experiment 
(OGLE, \citealt{ogleref}) collaboration alerted
the microlensing community to OGLE-2015-BLG-1459 via its Early Warning system 
\citep{ews1}\footnote{http://ogle.astrouw.edu.pl/ogle4/ews/ews.html/}
based on observations
from their 1.3m telescope in Chile, which were carried out with
cadence $\Gamma=3\,{\rm hr}^{-1}$, primarily in $I$ band. The event
lies at (RA,DEC) = (18:00:50.40, $-28$:40:15.7), corresponding to
$(l,b)=(1.92,-2.73)$.
From Chile,
the event appears to be a simple point lens (1L1S) event that peaks
relatively faint $I_{\rm peak}\sim 18$. However, the Korea
Microlensing Telescope Network (KMTNet, \citealt{kmtnet}) independently
observed this field from three different sites during their
commissioning-year observations. KMTNet also observed primarily in $I$
band, with cadence $\Gamma=7\,{\rm hr}^{-1}$ from each of its three
1.6m telescopes in Chile (KMTC), South Africa (KMTS)\footnote{KMTS was
down for engineering during all but the tail end of the event and so
the data are not used here.}, and Australia (KMTA). KMTA
data showed a strong anomaly just after the peak.  See Figure~\ref{fig:lc}. 
{The OGLE and KMTNet data were reduced using their difference image
  analysis \citep[DIA; ][]{alard98} photometry pipelines
  (\citealt{wozniak2000} and \citealt{albrow09}, respectively).}

The Microlensing Observations in Astrophysics (MOA) collaboration
also took data of this field using their 1.8m telescope at Mt.\ John,
New Zealand, which employs a broad $R/I$ filter.  These data were originally
believed to be too low quality to be of use and were not initially
reduced.  And indeed, even after careful re-reduction {using a variant of DIA \citep{bond17}}, they do not 
significantly constrain any of the geometric parameters.  However,
because they are in a slightly different passband, they do constrain
the flux parameters, which proves crucial to resolving the degeneracy
between the models (Section~\ref{sec:resolve}).

{It is well known that the photometric errors that come from the
  photometry reduction algorithms are imperfect and can underestimate
  the true errors. Therefore, we rescale the error bars for each
  dataset using a variant of the method in \citet{mb11293}, which
  implicitly assumes that the underlying errors follow a gaussian
  distribution. For this event, the vast majority of the data are
  taken when the event is below sky, but the data on the night of the
  anomaly are above sky. We cannot assume that the photometric
  algorthims adequately account for this transition. Therefore, we
  renormalize the error bars on the peak night (HJD$^{\prime} = 7200$)
  separately from the data for the rest of the event such that the
  total $\chi^2$ per degree of freedom is 1. The error renomalization
  factors are given in Table \ref{tab:errorbars}.}

\section{Analysis}
\label{sec:anal}

\begin{figure}
\centering
\includegraphics[width=\textwidth]{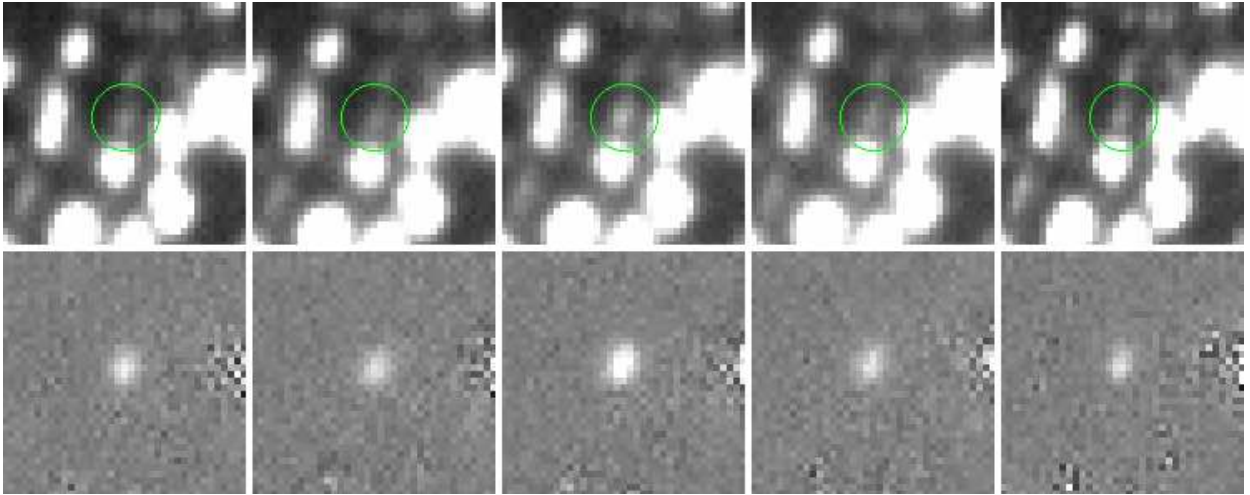}
\caption{\noindent {Actual (upper panels) and subtacted (lower panels)
  images from KMTA for the five observations centered on the ``outlying point'' (i.e., for HJD$^{\prime}\equiv$ HJD$-$2457200.0 = 0.188,
  0.194, 0.200, 0.207, and 0.213, left to right).}}
\label{fig:images}
\end{figure}

The anomaly (Figure~\ref{fig:lc}) consists of two features: a broad
bump and an outlying point at HJD$^{\prime}=7200.200$, which is $\sim
0.45$ magnitudes (a factor of $\sim 1.5$ in flux) brighter than the
two neighboring points taken $\sim 9\,$min earlier and
later. {Figure~\ref{fig:images} shows the observations and subtracted
images at these epochs.} As we show below, each of these features is
subject to two interpretations, i.e., each can indicate the presence
of an additional source or an additional lens component.

\subsection{{Models with 3 Bodies}}

To facilitate the analysis, we begin by
fitting just the OGLE and KMTNet data and by temporarily removing the
outlying point. {That is, we fit for models with a 2-body lens and
  a single source star (2L1S) and models with a single lens and two
  source stars (1L2S).}
 
{We first fit the pruned light curve with a 2L1S model.}
Such
models have 3 parameters describing the underlying stellar event (the
impact parameter, $u_0\theta_\e$, the time of closest approach, $t_0$,
the Einstein crossing time, $t_\e$), 3 parameters describing the
planet (its mass ratio, $q$, its projected separation, $s\theta_\e$, 
and the angle between
the source trajectory and the planet-star axis, $\alpha$), and the
normalized source radius $\rho=\theta_*/\theta_\e$. A thorough search
of parameter space yields two solutions (``close'' and ``wide'', {$s<1$ and $s>1$, respectively}), which are
related by the well known $s\leftrightarrow s^{-1}$ degeneracy for
central caustics \citep{griest98,dominik99}.

We then fit the same light curve to a single lens that microlenses
a binary source (1L2S).  The minimum requirement for such a fit,
which we employ here, is five geometric parameters: $t_\e$, plus
two pairs of $(t_0,u_0)_i$, one for each source, $i=1,2$.  In addition,
this model requires an extra flux parameter $q_F$, which is the ratio
of source fluxes in $I$ band (which is then treated as being the
same at all observatories).

Figure~\ref{fig:lcbin} shows the light curve in the region of the
cusp approach together with these two models, i.e., the 1L2S model
and the close 2L1S models (the wide model being almost identical).  Both
models explain the broad bump in the
anomaly. However, neither can explain the outlying point.

\subsection{{Models with 4 Bodies}}

{Although in principle, a single outlier could be due to a number
  of different phenomena, Occam's razor tells us it is far more likely
  that this is due to an additional microlensing effect.} There are
three ways to modify these models to account for the outlying
point{, all of which require adding an additional body to the
  model}.

In the first of these, one adds an additional source to the
1L2S model to obtain 1L3S.  The additional source is positioned
so that it is transited by the lens at the time of the spike.
The extremely short duration of the spike (combined with its modest
peak amplitude) is then attributed to the extreme faintness of
this third source.  In
fact, both additional sources must be quite faint, a point
to which we return below.  This minimalist version of the 1L3S
model requires three more geometric parameters $(t_0,u_0,\rho)_3$, as well
as an additional flux ratio parameter $q_{F,2}$.

\begin{figure}
\centering
\includegraphics[width=16cm]{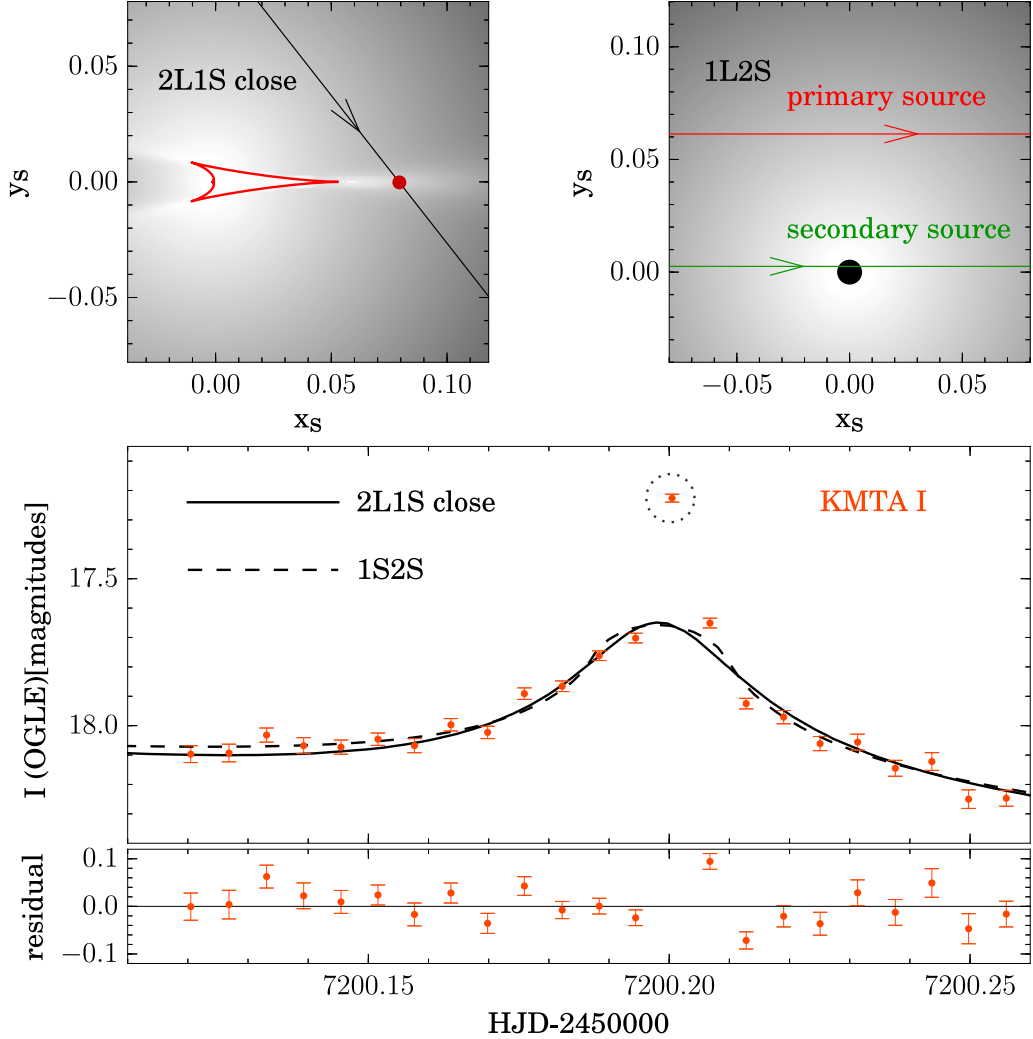}
\caption{\noindent Best-fit binary-lens (2L1S) and binary
source (1L2S) models to all the data
shown in Figure~\ref{fig:lc} except that the high point at
7200.200 (dotted circle) is excluded.  Both models are reasonably good, with
the main deviation in KMTA data explained either by a typical
4-pronged ``central caustic'' due to a planet (2L1S) or a second
source that is 1.7 mag fainter than the primary but lies much
closer to the path of the source.  There are
two degenerate 2L1S models, which yield very similar
light curves.  Hence, only the ``close'' solution is shown.
}
\label{fig:lcbin}
\end{figure}

In the second model, one adds a second source to the 2L1S model to
produce 2L2S.  The spike is then explained by the second source
passing over the cusp at almost exactly the same time (within 2 minutes)
that the first source passes over the ``magnification spike'' that
extends away from the tip of the caustic shown in Figure~\ref{fig:lcbin}.
This model requires the same three additional parameters as the one
described in the previous paragraph.

\begin{figure}
\centering
\includegraphics[width=16cm]{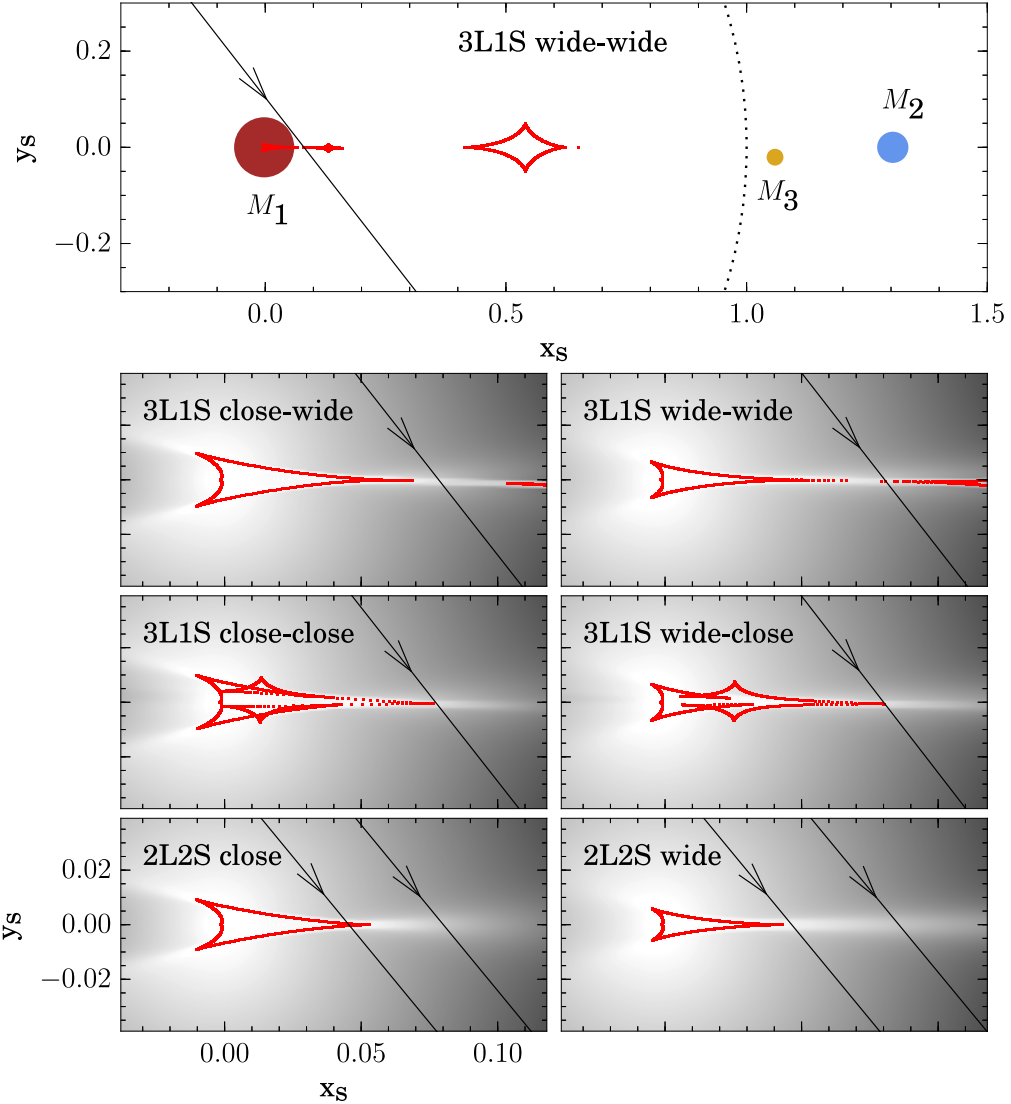}
\caption{\noindent 
Caustic geometries of the four 3L1S solutions are shown
in the middle four panels.  In each case, the high point at HJD 7200.200
is explained by the presence of a third body that, because it
is roughly aligned with the planet-star axis, ``extends'' the
caustic caused by the planet in Figure.~\ref{fig:lcbin}.  Upper
panel is a zoom-out of the wide-wide solution, showing the
full geometry.  In the bottom two panels, the two 2L2S solutions
are shown.  In these cases, the high point at HJD 7200.200 is
explained by a second source passing over the cusp seen in the upper-left
panel of Figure~\ref{fig:lcbin}.
}
\label{fig:geom}
\end{figure}

In the third model, one adds a third lens to the 2L1S model to 
produce 3L1S.  In this case, the spike is explained by the
third body distorting (lengthening) the cusp seen in Figure~\ref{fig:lcbin}
so that the source passes directly over it.  This requires three
additional geometric parameters relative to the 2L1S model, i.e.,
an additional pair of $(s,q)_2$ and an angle $\psi$ between the binary
axis and the line connecting the third body to the primary.
In this case there are four degenerate solutions, i.e., a close-wide
degeneracy \citep{griest98,ob120026} for each of the two low-mass companions.
See Figure~\ref{fig:geom}.

{Note that this does not represent an exhaustive search for triple
    lens solutions \citep[see][for an example of such a
    search]{ob160613}. However, Figure \ref{fig:3L1S} shows that these models
    explain all of the major features of the light curve. While there
    may be other 3L1S models that also fit the data, there is no
    indication (e.g., via significant residuals) that an alternative
    triple lens model would give an improved fit. We will revisit to
    this point in Section \ref{sec:resolve}.}

As shown in Figure~\ref{fig:lc} and quantified in 
Table~\ref{tab:chi2}, each of the
three of these basic models have variants that
fit the data approximately equally well.  That is, 
{the $\chi^2$ of the ``xallarap'' variant of the 1L3S model 
(see Section~\ref{sec:1l3s}) is comparable to the $\chi^2$ of} the better of the two 2L2S models and
the best of the four 3L1S models.  
Note that we do not show
the parameter values for the various models listed in Table~\ref{tab:chi2}
because these are very similar to the ones 
that we discuss in Section~\ref{sec:resolve}.

We first investigate the physical properties of these three models in
Section~\ref{sec:phys}. {Then, in Section~\ref{sec:resolve} we ask
  how we can distinguish between the three models given that the
  formal statistical difference between the models compared to OGLE and KMTNet
  data is insignificant.}

\section{Physical Properties}
\label{sec:phys}

\subsection{1L3S}
\label{sec:1l3s}

\begin{figure}
\centering
\includegraphics[width=16cm]{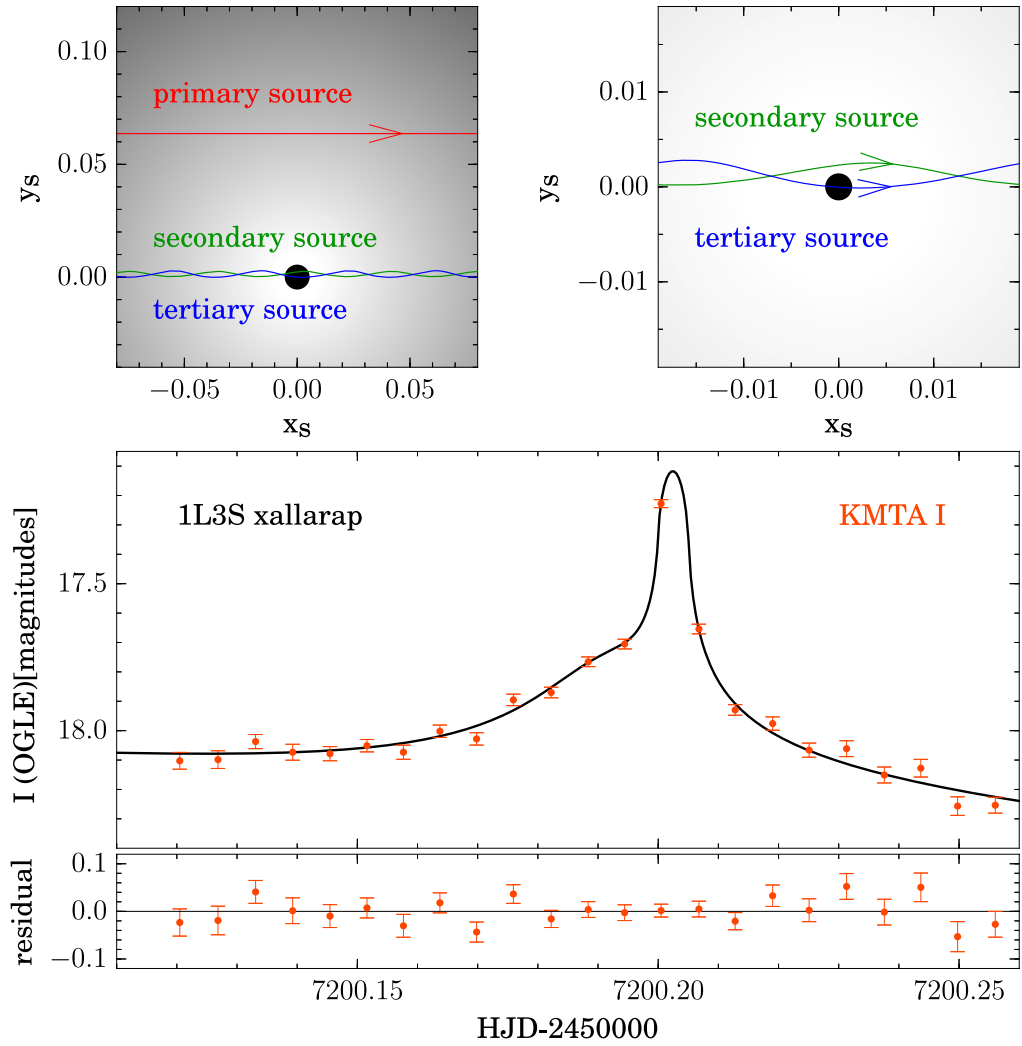}
\caption{\noindent 
Light curve and geometry for ``xallarap'' version of 1L3S,
in which the secondary and tertiary sources are modeled as being in
a circular orbit 4.5 hour orbit.  
}
\label{fig:xal}
\end{figure}

The sources of the 1L3S solution {likely form a
  gravitationally-bound hierarchical triple}.
In the static version of this solution, the primary is responsible
for the ``main event'', and the two fainter sources are each responsible
for one of the anomaly features.  As mentioned in Section~\ref{sec:anal},
the two fainter sources induce peaks at almost exactly the same time,
implying that they are projected close to each other within the Einstein ring
$\Delta u=0.0031$, where $\Delta {\bf u}\equiv 
[(t_{0,2}-t_{0,3})/t_\e,u_{0,2}-u_{0,3}]$.  See Table~\ref{tab:1l3s}, below.
This is actually only  a few times
larger than the normalized source size $\rho=0.0006$, meaning that these
two stars form an extremely tight binary (unless they are seen in
an extremely unlikely chance projection).  This in turn implies
that treating this binary as static is not a reasonable approximation,
since the two components are likely moving at several hundred kilometers
per second.  We therefore fit them to a standard xallarap (binary-source
motion) model, which allows for circular motion characterized by
four parameters, i.e.,
the 3-D separation $\chi_\e$ (scaled to $\theta_\e$),
and an arbitrary inclination, orientation, and phase of the orbit.  
See Figure~\ref{fig:xal} as well as Table~\ref{tab:1l3s_xa}, below.  
To determine the period, we assume
a total binary mass of $0.5\,M_\odot$ and a tertiary source radius
$R_3 = 0.2\,R_\odot$ (see below), adopt $\rho_3=0.0006$, and then apply Kepler's
Third Law.
This model hardly
improves $\chi^2$, but the parameters are quite reasonable. It is
also the case that most (perhaps all) very close binaries are in
hierarchical triples, which may point to a Lidov-Kozai origin
\citep{fabrycky07}.  Thus, this solution is in every respect quite
reasonable.

Its only ``bizarre'' feature is that both sources in the compact
binary are quite faint, lying $\Delta I_2=9$ and $\Delta I_3=10$
below the clump of the color magnitude diagram, where
$\Delta I_i\equiv I_{s,i}-I_{\rm clump}$ and $I_{\rm clump}$ is the
magnitude of the clump.  These would be the two faintest sources
ever reliably measured in a microlensing event.  However, given that
we see two short-lived perturbations peaking at nearly the same
time, it is not particularly surprising that they would be due
to a close, faint binary.  If the sources were brighter,
their flux would have been detectable over a broader portion of the light
curve.

\subsection{2L2S}
\label{sec:2l2s}

In terms of the physical characteristics of the lens system, the
2L2S and 3L1S are very similar, except that the latter
contains an additional, very low-mass object.  See Tables~\ref{tab:2l2s}
and \ref{tab:3l1s}, below. 
In both cases
the mass ratio of the (principal) planet to the host is 
$q\sim 3\times 10^{-3}$ and the lens-source relative proper motion
is quite high, $\mu\sim 50\,\masyr$.  This high proper motion
implies a very nearby lens, which (given the inverse relation between
$M$ and $\pi_\rel$ at fixed $\theta_\e$ specified by 
Equation~(\ref{eqn:thetaedef})), implies a very low host mass.
In both cases, this leads to a brown-dwarf host orbited by a sub-Saturn
(probably Neptune-class) planet.
We follow through this reasoning in some detail for 2L2S and then
briefly recount the minor differences for 3L1S.

In all solutions (1L3S, 2L2S, 3L1S) the (total) source flux  is about
$f_{s,\rm ogle}\simeq 0.055$ (normalized to $I=18$) and the mean color of the
source(s) (determined from regression) is $\Delta(V-I)=-0.14$ mag blueward
of the clump.  For the multi-source solutions, we lack separate
measurements for the colors of the different sources because the
secondary (and possibly tertiary) do not contribute substantially
to the total flux during times when there are $V$ data.  Nevertheless,
since the total flux lies $\sim 4.5$ mag below the clump, and is relatively
red, we can assume that both (or all three) of the sources lie on a single
main-sequence isochrone.  We approximate the color magnitude relation
on the main sequence as $\Delta M_I = 2.4\Delta(V-I)_0$.  Then, from the
measured flux ratio of the 2L2S model $q_F=0.189$, we can infer
that the secondary source is $\Delta[I,(V-I)]=(6.5,0.51)$ fainter
and redder than the clump.

It is the secondary source that is important in 2L2S because this
is the source that transits the cusp, and for which there is a normalized
source radius measurement, $\rho_2 = 0.73\pm 0.13$.  We combine the
above estimate of the source position relative to the clump with
the dereddened clump centroid $[(V-I),I]_{0,\rm clump}=(1.06,14.38)$
\citep{bensby13,nataf13}, the $VIK$ color-color relation of \citet{bb88}
and the color/surface-brightness relation of \citet{kervella04}, to
obtain $\theta_* = 0.48\,\mu$as \citep{ob03262}.  This implies
\begin{equation}
\theta_\e = {\theta_*\over\rho} = 0.67\pm 0.12\,\mas
\qquad
\mu = {\theta_\e\over t_\e} = 48\pm 9\,\masyr
\label{eqn:thetaemu2l2s}
\end{equation}

The proper motion implies that the lens must be nearby.  For example,
the lens cannot be in the Galactic bulge at $D_L\sim 8\,\kpc$ because
then the lens-source relative velocity would be $\mu_\rel D_L \sim
1800\,\kms$, which would imply that either the lens or source was not
bound to the Galaxy. Therefore, it must be relatively close to the
Sun. Even if the lens is from the quite sparse Galactic halo
population, with typical transverse speed of $v_\perp\sim 200\,\kms$,
it lies at $D_L=v_\perp/\mu_\rel\sim 0.9\,\kpc$.  Stars in the thick
disk or thin disk populations are 50 or 500 times more common than
halo stars, but typically have $v_\perp\sim 100\,\kms$ or $v_\perp\sim
50\,\kms$, meaning the lens would be a factor 2 or 4 closer.  Since
the volume of available lenses scales $\propto D_L^3$ (thus $\propto
v_\perp^3$), the disk and thick disk scenarios are about equally
likely, and the halo scenario is less likely than the combination of
the disk scenarios by a factor of $\sim 4$.  We normalize our analysis
to the thick disk scenario, keeping in mind that the masses and
distance could be higher or lower by a factor $\sim 2$, depending on
lens population.

A nearby lens in turn implies a low-mass lens. From
Equation~(\ref{eqn:thetaedef}), the total lens mass (essentially the
host mass) is $M=\theta_\e^2/\kappa\pi_\rel\simeq \theta_\e^2
v_\perp/\kappa\mu_\rel \simeq 0.024\,M_\odot(v_\perp/100\,\kms)$.
Thus, regardless of which population the lens system lies in (halo,
thick disk, or thin disk), the host is a low-mass brown dwarf (BD).
The planet therefore has mass $m_p = qM \sim 20\,M_\oplus$, i.e., slightly
heavier than Neptune.  Again, we should keep in mind that this value could be
a factor 2 higher or lower if the lens lay in the halo or thin disk,
respectively.

{In this section, we have performed the calculations for the
  ``close'' solution given in Tables~\ref{tab:chi2} and \ref{tab:2l2s}
  because the ``wide'' solutions is disfavored by
  $\Delta\chi^2=15$. The ``wide'' solution gives qualtitatively
  similar answers.} We note, however, that the wide solution has a
proper motion that is larger by a factor 1.4 and so is even more
extreme.

\subsection{3L1S}
\label{sec:3l1s}

The host-planet system in 3L1S is very similar to the one in 2L2S.
As seen from Table~\ref{tab:chi2}, all four solutions have qualitatively
similar $\chi^2$ values.  We
trace the calculation for the wide-wide solution, which has the
best $\chi^2$.  

For 3L1S, there is of course only one source.  We find $\theta_*=0.73\,\mu$as.
Combined with $\rho=0.84\pm0.14\times 10^{-3}$, this yields
$\theta_\e =0.87\pm 0.15\,\mas$, and $\mu=68\pm 13\,\masyr$.
Normalizing to the ``thick disk'' ($v_\perp= 100\,\kms$) case,
we find $D_L\sim 0.3\,\kpc$ and $M=0.021\,M_\odot$, and $m_p=qM=21\,M_\oplus$.
That is, very similar to 2L2S.

However, for 3L1S, there is also a third body with mass (normalized
again to the ``thick disk'' case), $m_m = 0.15\,M_\oplus$.  From 
Figure~\ref{fig:geom},  this third mass lies projected close to the
second, and so could possibly be its ``moon'', with mass ratio
$q_3/q_2=0.0076$, i.e., about a factor 1.6 smaller than the Moon/Earth mass
ratio.  We discuss the issues related to such an inference 
in Section~\ref{sec:discuss}.

\section{Resolution of the Degeneracy}
\label{sec:resolve}

The degeneracy reported here is basically a ``multiplicity'' of
the one first pointed out by \citet{gaudi98} between planetary
2L1S solutions and 1L2S solutions, particularly those with a
close source approach to a faint secondary.  The most 
secure way to distinguish between these two interpretations
is to measure the color difference between the two (putative) sources.
Since microlensing events involving a single source are basically
achromatic\footnote{with a very modest exception when the source is resolved
during a caustic crossing}, an evolution of apparent source color 
during an event is an ironclad indicator of multiple sources.

Of course if the two (or multiple) sources happen to have the same
color, then their combined, magnified light will also have this color.
However, in the present case, the secondary is two magnitudes
fainter than the primary in the 2L2S case, and the two sources
are 4.5 and 5.5 magnitudes fainter than the primary in the 1L3S case.
{The primary lies $\Delta I \sim 4.5$ magnitudes below the clump, making it} a fairly red (probably unevolved)
main-sequence star. {Given the observed flux ratios $q_{F,2}$ and $q_{F,3}$ this in turn implies, a rough} color offset
$\Delta(V-I)_2= -2.5\log(0.189)/2.4=0.75$ between the primary and secondary
for 2L2S, and $\Delta(V-I)_2=1.9$, $\Delta(V-I)_3=2.3$ for the secondary
and tertiary in 1L3S.

Both the OGLE and KMTNet surveys routinely take $V$-band measurements.
However, since the fundamental purpose of these measurements is to 
measure the source color (primarily in order to determine $\theta_*$),
the cadence of these observations is set to obtain a few magnified
points for the case of a relatively ``short'' event
(which might be a few hours to a few days depending on field
being observed).  As a result, in
2015, KMTNet obtained 1/6 points in $V$ band from KMTC and no
in $V$ band from either KMTS or KMTA.  Hence{, since the anomaly was only observable by KMTA,} there were no
$V$ data that could probe the color of the anomalous part of the
light curve.

However, incorporating MOA data can potentially yield 
the necessary color information.  The difference between MOA's broad $R/I$ 
filter ($R_{\rm moa}$) and standard
$I$ band (used by KMTNet) is much smaller than the difference between
$V$ and $I$.  The exact value is field dependent, but for example
\citet{gould10} found that for the field of MOA-2007-BLG-192, the
difference was $\Delta(R_{\rm moa} -I) = 0.26\Delta(V -I)$.    Thus,
we expect that if 1L3S is the correct model, and in the approximation
that the flux normalization of the MOA data is completely set
by the portions of the light curve away from the anomaly (where
the total flux is dominated by the primary source, which is
relatively blue) then the MOA light curve should be substantially
fainter than the KMTA light curve in the region of the anomaly.
On the other hand, if 3L1S is correct, then we expect that the
two light curves should be everywhere comparable. 
{This would be true of any 3L1S model, so this test could rule out
  all 3L1S models, even if we have not found all possible 3L1S
  solutions.}

The case of 2L2S should be qualitatively similar to 3L1S because
the excess light for the main anomaly is due to the primary
source passing close to a caustic.  Hence this main-anomaly
region should have basically the same color as the overall
light curve.  While we do expect the color to turn redder in the immediate
neighborhood of the spike, where the secondary contributes of
order half the light, unfortunately MOA does not have data
during this spike.

\begin{figure}
\centering
\includegraphics[width=16cm]{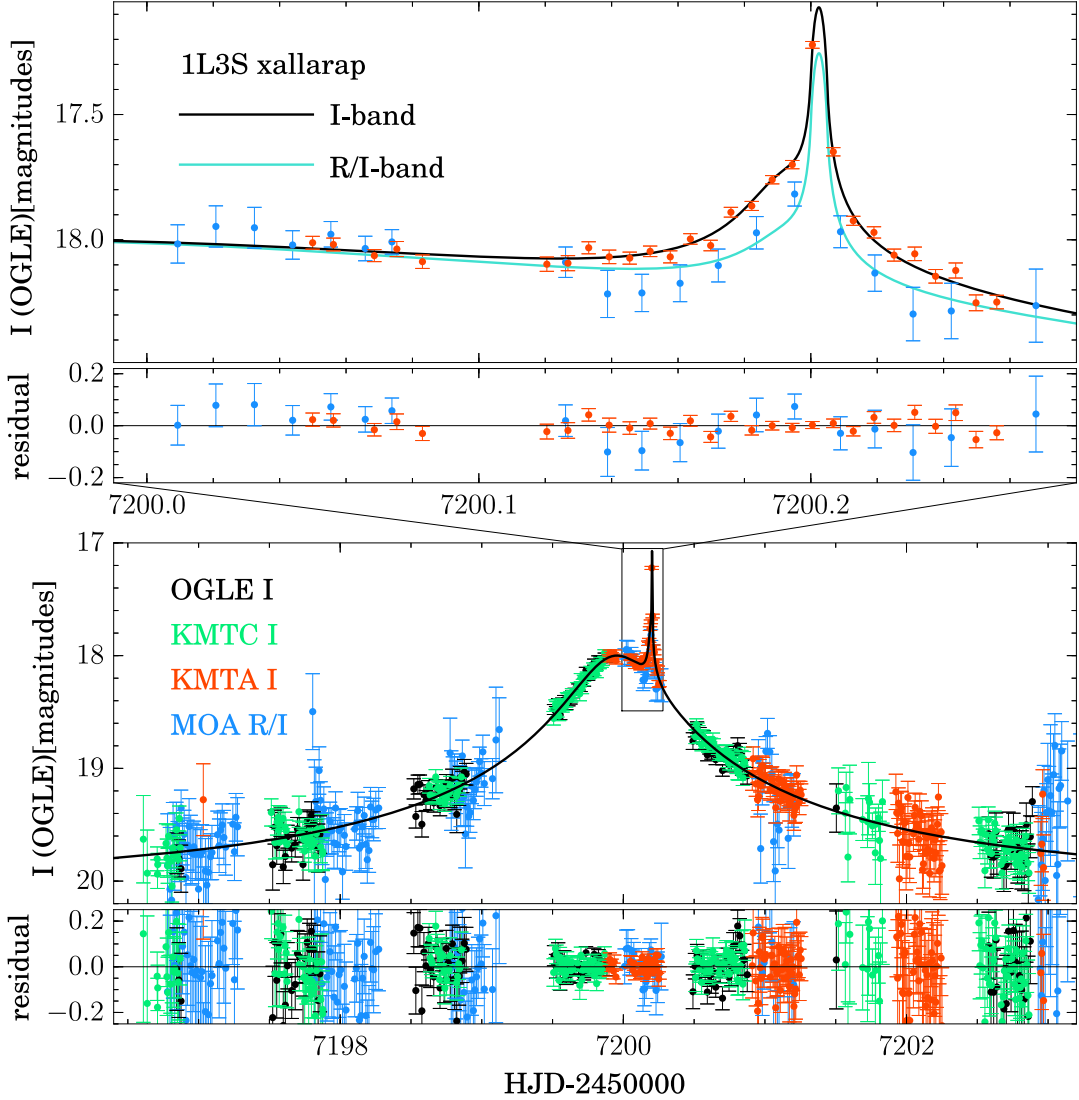}
\caption{\noindent 
Joint fit of OGLE, KMTNet, and MOA data for the 1L3S xallarap solution.
Because the MOA $(R/I)$ passband is bluer than the OGLE and KMTNet $I$-band,
there is an additional photometric degree of freedom, $q_{FR}$, the flux ratio
of the second and third sources (assumed the same) to the primary.
In the region where these sources dominate, near 7200.2, the MOA data
fall below the KMTA data, as one would expect if this were the correct
model.  In principle, one would expect the MOA data to also fall below
KMTA earlier in the night, near 7200.05, where the secondary and tertiary
sources play a significant role.  However, the quality of the MOA
data in other magnified portions of the light curve (lower panel)
is too noisy to permit strict alignment of the flux scale based
on this part of the light curve alone.  Hence, most of the
test comes from comparison of the early part of the anomalous night
relative to the later part.
}
\label{fig:1L3S}
\end{figure}

\begin{figure}
\centering
\includegraphics[width=16cm]{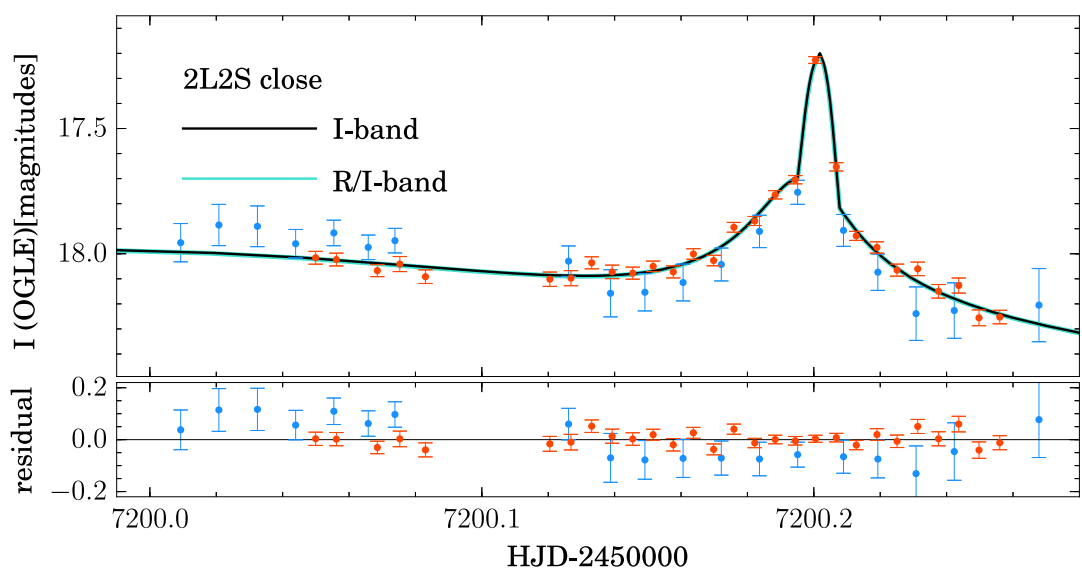}
\caption{\noindent 
Joint fit of OGLE, KMTNet, and MOA data for the 2L2S solution.
The offset of the MOA and KMTA points at the beginning relative
to the end of the night is, of course, the same as in Figure~\ref{fig:1L3S}.
However, the model cannot account for this as well because the secondary
only dominates the light in the immediate neighborhood of the spike,
where there are few MOA data points.  (The zoom-out of the full light curve
is not shown because it is indistinguishable from the lower panel of
Figure~\ref{fig:1L3S}.)
}
\label{fig:2L2S}
\end{figure}

\begin{figure}
\centering
\includegraphics[width=16cm]{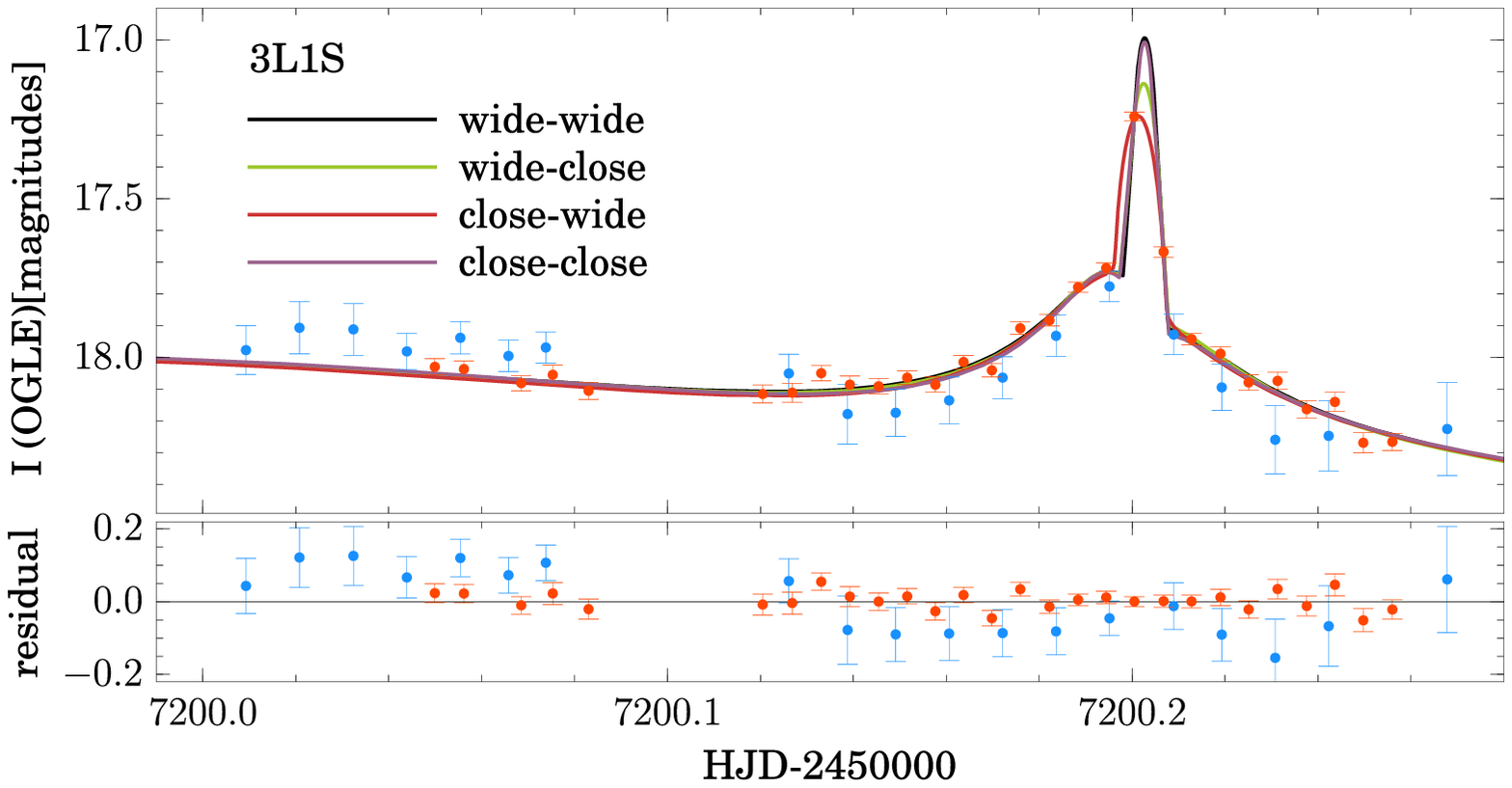}
\caption{\noindent 
Joint fit of OGLE, KMTNet, and MOA data for the 3L1S solution.
In this case there is only one source, so the MOA data should
should track the KMTA data.  Hence, there is no way to accommodate,
within the model, the fact that they are brighter in the beginning
of the night relative to the end.   (The zoom-out of the full light curve
is not shown because it is indistinguishable from the lower panel of
Figure~\ref{fig:1L3S}.)
}
\label{fig:3L1S}
\end{figure}

{Thus, we have a strong test that can distinguish the 1L3S solution from the 2L2S and 3L1S solutions: either the MOA data will show a color-dependent effect or they will not.} With these predictions in mind, we incorporate MOA data into all
fits, with results shown in Tables~\ref{tab:1l3s}--\ref{tab:3l1s} and
Figures~\ref{fig:1L3S}--\ref{fig:3L1S}.  We note that for the 1L3S
solutions, we assume that the two fainter sources have the same
flux.  This is because (in contrast to the $I$-band data) the
MOA data do not cover the short spike at HJD$^\prime= 7200.20$, and
so do not distinguish between the two fainter sources.  Moreover, for the
2L2S solutions, we impose $q_{FR}<q_F$, as discussed below.

Figure~\ref{fig:1L3S} is in agreement with the main prediction of the
1L3S model: the MOA data lie significantly below the KMTA data during
the entire latter part of the night, as expected, when the two faint,
red sources contribute a major part of the total flux.  According to
the above predictions one would also expect that the MOA data would
lie below the KMTA data during the first half of the peak night,
albeit by substantially less, because the two faint, red sources
contribute somewhat to the total light in this portion.  Instead, the
MOA data are coincident or slightly above the KMTA data.  However,
this discrepancy can be explained by the relatively noisy character of
the MOA data on non-peak nights. {This leaves some freedom for the
  model to adjust the primary source flux to better fit the data over the
  anomaly. If the data were better, one would expect the primary source flux to be entirely set by} the $I$-band dominated model
during the epochs when the (relatively blue) primary completely dominates
the light curve.  Given this, the test really only predicts that
the MOA data will lie further below KMTA data at the end of the night
than in the beginning.

Figures~\ref{fig:2L2S} and \ref{fig:3L1S} are also in accord with the
predictions of the 1L3S model and contradict, respectively, the 2L2S and
3L1S models.  First, as predicted, these two figures look qualitatively
similar.  Second the models for MOA data track the $I$-band models,
while the MOA data are systematically higher than the model at the 
beginning of the night and lower at the end.  That is, the models
have no way to accommodate the observed change in color from the
beginning to the end of the night.  We note that the models ``try''
to accommodate this change by making the secondary bluer than the
primary in the 2L2S model, i.e., $q_{FR}>q_F$.  Since the secondary
is 2 mag fainter than the primary and both are on the main sequence,
we prohibit this unphysical tendency by imposing a boundary in the chains.

Comparing the best $\chi^2$ values for each solution (two for 1L3S,
two for 2L2S, four for 3L1S), we see that the 1L3S solution is
preferred 
by $\Delta\chi^2=7.5$ for 2 dof over 2L2S and 
by $\Delta\chi^2=11.9$ for 4 dof over 3L1S.
These have formal probabilities of $\exp(-\Delta\chi^2/2)=2.3\%$
and $(\Delta\chi^2/2 +1)\exp(-\Delta\chi^2/2)=3.4\%$, respectively.

Even though these p-values are not extremely low, we consider
the degeneracy to be clearly resolved in favor of 1L3S.  This
work began by investigating OGLE and KMTNet data because these
were all that appeared to be available.  We then made special efforts
to recover the MOA data, solely to test whether there was color
evolution as predicted by one of the degenerate models and not
by the others.  Hence, since we have asked a very simple, one parameter
question of the MOA data, we consider it reasonable 
that the above p-values should
be taken at face value.

\section{Discussion}
\label{sec:discuss}

The analysis of OGLE-2015-BLG-1459 lays bare both the promise
and the challenges of exo-moon research using microlensing.
On the one hand, it serves as a proof of concept: if there
had been a 3-body, BD-host/Neptune/Mars lens system present,
we would have detected it.  Moreover, we would have been able
to demonstrate that the 3L1S and 2L2S solutions were preferred
over the 1L3S solution.  This would have left an ambiguity
between 3L1S and 2L2S, but as we briefly mention below, this
could have been resolved by followup spectroscopy.

On the other hand, this event also illustrates two major difficulties
confronting microlensing exo-moon studies, one that is practical and the
other that is of a more fundamental character.

{The practical problem is that microlensing experiments do not take
  alternate band (usually $V$-band) data often enough to measure a
  color change for an exomoon. Measuring the color of a ``short''
  event requires taking data in two different bands during that
  event. Exactly what is meant by ``short'' varies depending on the
  application, and may be as short as a few hours for the planets
  targeted by current surveys. However, in the case of this exo-moon
  candidate, ``short'' corresponds to a single data point. If the rest
  of the light curve had proved achromatic, a color could only have
  been reliably measured by alternating bands between
  observations. Such observations would be required to distinguish
  between the 2L2S and 3L1S solutions (had they been viable).}

{However, from the standpoint of a microlensing experiment focused
  on finding planets rather than moons (i.e., all current
  experiments), alternating observations between two bands would be
  extremely wasteful. These experiments take of order $10^{12}$
  photometric measurements per year. While there may be some real
  microlensing events that occur on timescales comparable to the
  cadence (i.e., consisting of only a single point), there is no way
  to identify them amongst the enormous number of cosmic ray events
  and other image artifacts that occur on the same
  timescales. Therefore, there is no need to measure their colors, and
  so no need for a second band of observations on that timescale.}
Moreover, given the high reddening in typical microlensing
fields, typical sources yield 5--10 times fewer photons in $V$ than $I$
for the same exposure time.  Hence, attempting to get one $V$ for
each $I$ measurement would greatly undermine the overall experiment.

{In contrast}, there are only a few dozen microlensing planets discovered
per year and each planet is typically characterized by a few dozen
data points.  Hence, there are only of order $10^3$ points that
could potentially be sensitive to exo-moons.  The handful of these
points that show a potential signatures can easily be vetted by examining
the images (as we in fact did in this case).  Hence, from the standpoint of
finding exo-moons, equally dense $V$ and $I$ measurements would not
be at all wasteful.

One possible compromise would be to have intensive $V$-band measurements
only in a small fraction of the sky-area covered, in particular
the area with the highest number of events.  For example, KMTNet
currently spends 1/4 of its time on the highest density field
(BLG02 + BLG42). If this field were covered $V$:$I$ as 1:1 or 1:2, then
this would reduce the overall cadence of the experiment by factors
of 4/5 or 8/9, respectively.  This might be an acceptable cost
for probing new parameter space.

The second challenge is more fundamental.
If the 3L1S solution had been the correct interpretation, then since
the two smaller bodies are projected close to each other on the sky,
the Mars-class body could have been a moon of the Neptune-class body.
If it were a moon, then there are exactly three things that can be
said about this planet-moon system based on the microlensing light
curve.  First, their host would have been a low-mass BD.  Second,
their mass ratio would have been about 100:1.  Third, the plane of the
planet-moon orbit would have been significantly misaligned from the
plane of BD-planet orbit. To be bound, the moon must lie in the
planet's Hill sphere, which has radius $a(q/3)^{1/3}$, where $a$ is
the semi-major axis.  Hence, if the orbits were co-planar, then a
bound orbit requires $(1 - s_2/s)(q/3)^{-1/3} <1$.  In fact, assuming
coplanar orbits, this ratio is 1.9 for the wide-wide case and higher
for all others.  However, if the planet-moon system were seen roughly
face-on while the star-planet axis was inclined by at least
$\theta>\tan^{-1}(1.9) > 62^\circ$ (as for the regular moons of
Uranus) then the system would satisfy this condition.

However, it also would have been possible that the third body was not
bound to the Neptune-class planet, and in fact independently orbited
the star on a wider or closer orbit. Because of the nature of the
microlensing technique, we can detect only the projected positions of
these two bodies and cannot generally tell whether they are in front
of or behind the plane of the lens. Therefore, it may be that we would
happen by chance to observe the system at a point in the orbits of two
planets that makes them appear to be close to each other in projection
even though they orbit the host independently. Unfortunately, there
is no way to distinguish between these possibilities based on the
microlensing data.  Nor would there be any possibility of further
investigating the system with present, or presently conceived,
instruments.

Moreover, there will always be this ambiguity even in cases
for which the third body lies projected within the Hill sphere
of the second.  It may seem more likely that two bodies projected close
together would be bound to each other in a planet-moon system: two
bound bodies will always appear close in projection because they are
physically close. The alternative requires that we have observed the
two planets at a special time in their orbits by chance. However, we
must also take into account observational bias. Even if the two
planetary bodies are not physically related, the probability of
detectable signals from both bodies is increased if they appear close
together in projection. Both bodies will preferentially be found close
to the star's Einstein radius, and the probability of detecting a
small third body will be enhanced by its proximity (in projection) to
the second body. Thus, the study of microlensing exo-moons must be
done on a statistical basis and will also require systematic
simulations to quantify this observational bias.  At the next level
of complexity, such simulations should take account of dynamical interactions,
such as Lidov-Kozai oscillations, which might affect the stability
of marginally Hill-stable systems.

These issues must be taken into account not only in existing
ground-based surveys like OGLE and KMTNet but also in the future {\it
  WFIRST} microlensing survey \citep{wfirst}.  {\it WFIRST}'s precise
photometric precision makes it potentially far more capable of
detecting the subtle signals from exo-moons as compared to present
surveys. This work suggests that such studies would benefit from
frequent data in a second band to distinguish additional lens planets
or moons from additional source stars. In addition, to more fully
characterize very short perturbations with 
$t_* \equiv \rho t_\e\lesssim 9\,$min
requires a faster cadence. This might be accomplished by a tiered
observing strategy, such as the one adopted by OGLE long ago in which
some fields are observed at a very high cadence at the expense of
other fields. In fact, in 2016 KMTNet changed its strategy to
follow the same principle. 

Finally, for completeness, we note that if the MOA data had strongly
preferred the (2L2S or 3L1S) solutions instead of 1L3S, then the
remaining ambiguity could have been resolved by followup spectroscopy.
For 2L2S, the secondary is only about 1.6 magnitudes fainter than the
primary and is likely moving at several tens of kilometers per second relative
to it.   Hence, it could be separately detected with an $R=20000$
spectrograph.  While the second source would be quite faint by
today's standards, $I\sim 23$, such spectroscopy would be in the
range of next generation (``30 meter'') class telescopes.

From OGLE-2015-BLG-1459, we can conclude that exo-moon studies with
microlensing will be challenging. Although we ultimately rejected the
exo-moon explanation for this event, the event provided the
first practical glimpse of what is required to meet these challenges.

\acknowledgments
Work by KHH was support by KASI grant 2017-1-830-03.
Work by WZ, YKJ, and AG were supported by AST-1516842 from the US NSF.
WZ, IGS, and AG were supported by JPL grant 1500811.  
This research has made use of the KMTNet system operated by the Korea
Astronomy and Space Science Institute (KASI) and the data were obtained at
three host sites of CTIO in Chile, SAAO in South Africa, and SSO in
Australia.
Work by C.H. was supported by the grant (2017R1A4A101517) of
National Research Foundation of Korea.
The OGLE Team thanks Prof. G. Pietrzy{\'n}ski for his contribution to the
collection of the OGLE photometric data.
The OGLE project has received funding from the National Science Centre, 
Poland, grant MAESTRO 2014/14/A/ST9/00121 to AU.
The MOA project is supported by JSPS KAKENHI Grant Number JSPS24253004, JSPS26247023, JSPS23340064, JSPS15H00781, and JP16H06287.

\input taberr

\input tabchi2

\input tab1L3S

\input tab1L3S_xa

\input tab2L2S

\input tab3L1S

\end{document}

%% file: author.tex
\author{\textsc{
K.-H. Hwang$^{1}$, 
A. Udalski$^{2}$,
I. A. Bond$^{3}$\\
\and 
M. D. Albrow$^{4}$, 
S.-J. Chung$^{1,5}$, 
A. Gould$^{1,6,7}$, 
C. Han$^{8}$, 
Y. K. Jung$^{1}$,
Y.-H. Ryu$^{1}$, 
I.-G. Shin$^{9}$, 
J. C. Yee$^{9}$, 
W. Zhu$^{6}$, 
S.-M. Cha$^{1,10}$, 
D.-J. Kim$^{1}$, 
H.-W. Kim$^{1}$, 
S.-L. Kim$^{1,5}$, 
C.-U. Lee$^{1,5}$,
D.-J. Lee$^{1}$,
Y. Lee$^{1,10}$, 
B.-G. Park$^{1,5}$, 
R. W. Pogge$^{6}$ \\
(KMTNet Collaboration)\\
M. Pawlak$^{2}$, 
R. Poleski$^{2,6}$,
M. K. Szyma\'{n}ski$^{2}$,
J. Skowron$^{2}$, 
I. Soszy\'{n}ski$^{2}$, 
P. Mr\'{o}z$^{2}$, 
S. Koz{\l}owski$^{2}$, 
P. Pietrukowicz$^{2}$, 
K. Ulaczyk$^{2}$\\
(OGLE Collaboration)\\
F. Abe$^{11}$, 
Y. Asakura$^{11}$, 
R. Barry$^{12}$,
D. P. Bennett$^{12}$, 
A. Bhattacharya$^{12}$,
M. Donachie$^{13}$, 
P. Evans$^{13}$, 
A. Fukui$^{14}$,
Y. Hirao$^{15}$, 
Y. Itow$^{11}$, 
K. Kawasaki$^{15}$,
N. Koshimoto$^{15}$, 
M. C. A. Li$^{13}$, 
C. H. Ling$^{3}$,
K. Masuda$^{11}$, 
Y. Matsubara$^{11}$, 
S. Miyazaki$^{15}$,
Y. Muraki$^{11}$, 
M. Nagakane$^{15}$, 
K. Ohnishi$^{16}$,
C. Ranc$^{12}$, 
N. J. Rattenbury$^{13}$, 
To. Saito$^{17}$,
A. Sharan$^{13}$, 
D. J. Sullivan$^{18}$, 
T. Sumi$^{15}$,
D. Suzuki$^{12,19}$, 
P. J. Tristram$^{20}$, 
T. Yamada$^{21}$,
T. Yamada$^{15}$, 
A. Yonehara$^{21}$\\
(MOA Collaboration)}}

\affil{$^{1}$Korea Astronomy and Space Science Institute, Daejon
34055, Korea}

\affil{$^{2}$Warsaw University Observatory, Al. Ujazdowskie 4,
00-478 Warszawa, Poland}

\affil{$^{3}$Institute of Natural and Mathematical Sciences, Massey
University, Auckland 0745, New Zealand}

\affil{$^{4}$University of Canterbury, Department of Physics and
Astronomy, Private Bag 4800, Christchurch 8020, New Zealand}

\affil{$^{5}$Korea University of Science and Technology, 
217 Gajeong-ro, Yuseong-gu, Daejeon 34113, Korea}

\affil{$^{6}$Department of Astronomy, Ohio State University, 140 W.
18th Ave., Columbus, OH 43210, USA}

\affil{$^{7}$Max-Planck-Institute for Astronomy, K\"{o}nigstuhl 17,
69117 Heidelberg, Germany}

\affil{$^{8}$Department of Physics, Chungbuk National University,
Cheongju 28644, Republic of Korea}

\affil{$^{9}$Harvard-Smithsonian Center for Astrophysics, 60 Garden Street, 
Cambridge, MA 02138, USA}

\affil{$^{10}$School of Space Research, Kyung Hee University,
Yongin, Kyeonggi 17104, Korea}

\affil{$^{11}$Institute for Space-Earth Environmental Research,
Nagoya University, Nagoya 464-8601, Japan}

\affil{$^{12}$Code 667, NASA Goddard Space Flight Center, Greenbelt,
MD 20771, USA; Email: david.bennett@nasa.gov}

\affil{$^{13}$Department of Physics, University of Auckland, Private
Bag 92019, Auckland, New Zealand}

\affil{$^{14}$Okayama Astrophysical Observatory, National
Astronomical Observatory of Japan, 3037-5 Honjo, Kamogata, Asakuchi,
Okayama 719-0232, Japan}

\affil{$^{15}$Department of Earth and Space Science, Graduate School
of Science, Osaka University, Toyonaka, Osaka 560-0043, Japan}

\affil{$^{16}$Nagano National College of Technology, Nagano
381-8550, Japan}

\affil{$^{17}$Tokyo Metropolitan College of Aeronautics, Tokyo
116-8523, Japan}

\affil{$^{18}$School of Chemical and Physical Sciences, Victoria
University, Wellington, New Zealand}

\affil{$^{19}$Institute of Space and Astronautical Science, Japan
Aerospace Exploration Agency, Kanagawa 252-5210, Japan}

\affil{$^{20}$University of Canterbury Mt.\ John Observatory, P.O.
Box 56, Lake Tekapo 8770, New Zealand}

\affil{$^{21}$Department of Physics, Faculty of Science, Kyoto
Sangyo University, 603-8555 Kyoto, Japan}

%% file: taberr.tex
\begin{table}[ht]
\begin{center}
\caption{{Parameters for Scaling Data Errorbars}\label{tab:errorbars}}
\begin{tabular}{lrr}
\hline
\hline
\multicolumn{1}{c}{Telescope} & 
\multicolumn{1}{c}{Number} & 
\multicolumn{1}{c}{k} \\
\hline
OGLE        & 2088 & 1.30\\
MOA         & 2186 & 1.09\\
MOA (peak)  &   19 & 0.98\\
KMTC        & 1399 & 1.63\\
KMTA        & 1091 & 2.61\\
KMTA (peak) &   34 & 1.50\\
\hline
\end{tabular}
\end{center}
\end{table}

%% file: tabchi2.tex
\begin{table}[ht]
\begin{center}
\caption{Comparison of $\chi^2$ for $I$-band models\label{tab:chi2}}
\begin{tabular}{llrrr}
\hline
\hline
\multicolumn{1}{c}{Model} & 
\multicolumn{1}{c}{Variant} & 
\multicolumn{1}{c}{$\chi^2$/dof} &
\multicolumn{1}{c}{$N_{params}$\tablenotemark{a}} &
\multicolumn{1}{c}{$N$ w/$\Delta I > 0.3$\tablenotemark{b}}\\
\hline
1L3S          & xallarap    & 4605.9/4598 & 13 & 438 \\
1L3S          & static      & 4608.2/4599 & 12 & 439 \\
\hline                                          
2L2S          & close       & 4603.8/4600 & 11 & 440 \\
2L2S          & wide        & 4618.7/4600 & 11 & 440 \\
\hline                                          
3L1S          & wide-wide   & 4604.7/4601 & 10 & 437 \\
3L1S          & wide-close  & 4604.8/4601 & 10 & 434 \\
3L1S          & close-close & 4608.2/4601 & 10 & 438 \\
3L1S          & close-wide  & 4608.3/4601 & 10 & 438 \\
\hline
\end{tabular}
\end{center}
\tablenotetext{a}{{The number of parameters of the model.}}
\tablenotetext{b}{{The number of data points $>0.3$ magnitudes above baseline.}}
\end{table}

%% file: tab1L3S.tex



\begin{table*}
\begin{center}
\caption{}
\begin{tabular}{lc}
\hline
\hline
\multicolumn{1}{c}{} & 
\multicolumn{1}{c}{1L3S static} \\
\hline
$\chi^2$/dof          & 6803.83/6803       \\
$t_0\ ({\rm HJD}')$     & 7199.946$\pm$0.004 \\
$u_0$                 &    0.065$\pm$0.005 \\
$t_{\rm E}$           &    4.921$\pm$0.291 \\
$\rho_*$              &    --              \\
$t_{0,2}\ ({\rm HJD}')$ & 7200.193$\pm$0.002 \\
$u_{0,2}\ (10^{-3})$    &    2.638$\pm$0.676 \\
$\rho_{*,2}\ (10^{-3})$ &    4.503$\pm$1.525 \\
$q_{F,2}$             &    0.014$\pm$0.003 \\
$t_{0,3}\ ({\rm HJD}')$ & 7200.202$\pm$0.001 \\
$u_{0,3}\ (10^{-3})$    &    0.281$\pm$0.165 \\
$\rho_{*,3}\ (10^{-3})$ &    0.631$\pm$0.229 \\
$q_{F,3}$             &    0.006$\pm$0.001 \\
$q_{F,R}$\ (MOA)         &    0.006$\pm$0.001 \\
$f_s$\ (OGLE)              &    0.056$\pm$0.004 \\
$f_b$\ (OGLE)              &    0.100$\pm$0.004 \\
\hline
\label{tab:1l3s}
\end{tabular}
\end{center}
\end{table*}

%% file: tab1L3S_xa.tex
\begin{table*}
\begin{center}
\caption{}
\begin{tabular}{lc}
\hline
\hline
\multicolumn{1}{c}{} & 
\multicolumn{1}{c}{1L3S xallarap} \\
\hline
$\chi^2$/dof              & 6800.96/6802       \\
$t_0\ ({\rm HJD}')$         & 7199.945$\pm$0.004 \\
$u_0$                     &    0.067$\pm$0.005 \\
$t_{\rm E}$               &    4.913$\pm$0.320 \\
$\rho_*$                  &    --              \\
$t_{0,(2,3)}\ ({\rm HJD}')$ & 7200.198$\pm$0.001 \\
$u_{0,(2,3)}\ (10^{-3})$    &    1.406$\pm$0.298 \\
$\rho_{*,2}\ (10^{-3})$     &    0.584$^{+1.150}_{-0.203}$ \\
$\rho_{*,3}\ (10^{-3})$     &    0.6\ (fixed)\\
$q_{F,2}$                 &    0.0145$\pm$0.0022 \\
$q_{F,3}$                 &    0.0067$\pm$0.0011 \\
$q_{F,R}$\ (MOA)             &    0.0056$\pm$0.0009 \\
$\chi_{E,X}$              &    0.0021$\pm$0.0008 \\
$\chi_{E,Y}$              &    0.0020$\pm$0.0006 \\
$\alpha$                  &    270.79$\pm$14.13  \\
$\delta$                  &    -32.11$\pm$29.98  \\
$f_s$\ (OGLE)                  &     0.056$\pm$0.005  \\
$f_b$\ (OGLE)                  &     0.100$\pm$0.005  \\
\hline
\label{tab:1l3s_xa}
\end{tabular}
\end{center}
\end{table*}

%% file: tab2L2S.tex
\begin{table*}
\begin{center}
\caption{}
\begin{tabular}{lcc}
\hline
\hline
\multicolumn{1}{c}{} & 
\multicolumn{2}{c}{2L2S} \\
\cline{2-3}
\multicolumn{1}{c}{} & 
\multicolumn{1}{c}{close} &
\multicolumn{1}{c}{wide} \\
\hline
$\chi^2$/dof          & 6808.45/6804       & 6825.04/6804       \\
$t_0\ ({\rm HJD}')$     & 7199.941$\pm$0.005 & 7199.914$\pm$0.009 \\
$u_0$                 &    0.069$\pm$0.003 &    0.066$\pm$0.004 \\
$t_{\rm E}$           &    4.613$\pm$0.201 &    5.106$\pm$0.231 \\
$s$                   &    0.857$\pm$0.009 &    1.260$\pm$0.014 \\
$q\ (10^{-3})$          &    2.374$\pm$0.207 &    2.068$\pm$0.209 \\
$\alpha$              &    0.890$\pm$0.008 &    0.875$\pm$0.009 \\
$\rho_*\ (10^{-3})$     &    0.688$\pm$0.595 &    1.795$\pm$0.871 \\
$t_{0,2}\ ({\rm HJD}')$ & 7200.057$\pm$0.010 & 7200.049$\pm$0.008 \\
$u_{0,2}$             &    0.039$\pm$0.003 &    0.036$\pm$0.003 \\
$\rho_{*,2}\ (10^{-3})$ &    0.542$\pm$0.161 &    0.457$\pm$0.164 \\
$q_{F}$               &    0.181$\pm$0.077 &    0.396$\pm$0.090 \\
$q_{F,R}$\ (MOA)         &    0.177$\pm$0.072 &    0.392$\pm$0.090 \\
$f_s$\ (OGLE)              &    0.061$\pm$0.003 &    0.054$\pm$0.003 \\
$f_b$\ (OGLE)              &    0.095$\pm$0.003 &    0.101$\pm$0.003 \\
\hline
\label{tab:2l2s}
\end{tabular}
\end{center}
\end{table*}

%% file: tab3L1S.tex
\begin{table*}
\begin{center}
\caption{}
\begin{tabular}{lcccc}
\hline
\hline
\multicolumn{1}{c}{} & 
\multicolumn{4}{c}{3L1S} \\
\cline{2-5}
\multicolumn{1}{c}{} & 
\multicolumn{1}{c}{close-close} &
\multicolumn{1}{c}{close-wide} &
\multicolumn{1}{c}{wide-close} &
\multicolumn{1}{c}{wide-wide} \\
\hline
$\chi^2$/dof      & 6817.60/6806       & 6817.83/6806       & 6813.16/6806       & 6812.87/6806       \\
$t_0\ ({\rm HJD}')$ & 7199.970$\pm$0.004 & 7199.967$\pm$0.004 & 7199.969$\pm$0.004 & 7199.970$\pm$0.004 \\
$u_0$             &    0.063$\pm$0.004 &    0.063$\pm$0.003 &    0.062$\pm$0.004 &    0.066$\pm$0.002 \\
$t_{\rm E}$       &    4.650$\pm$0.227 &    4.695$\pm$0.206 &    4.790$\pm$0.221 &    4.485$\pm$0.111 \\
$s$               &    0.845$\pm$0.010 &    0.840$\pm$0.010 &    1.294$\pm$0.013 &    1.302$\pm$0.012 \\
$q\ (10^{-3})$      &    2.508$\pm$0.222 &    2.611$\pm$0.228 &    2.667$\pm$0.238 &    2.959$\pm$0.225 \\
$\alpha$          &    0.906$\pm$0.009 &    0.906$\pm$0.008 &    0.910$\pm$0.008 &    0.916$\pm$0.007 \\
$\rho_*\ (10^{-3})$ &    0.552$\pm$0.123 &    0.895$\pm$0.123 &    0.683$\pm$0.118 &    0.617$\pm$0.129 \\
$s_2$             &    1.033$\pm$0.008 &    1.070$\pm$0.009 &    1.020$\pm$0.009 &    1.054$\pm$0.008 \\
$q_2\ (10^{-5})$    &    1.950$\pm$0.862 &    2.199$\pm$0.868 &    1.638$\pm$0.743 &    2.179$\pm$0.643 \\
$\psi$            &   -0.022$\pm$0.003 &   -0.020$\pm$0.003 &   -0.022$\pm$0.003 &   -0.023$\pm$0.003 \\
$f_s$\ (OGLE)          &    0.060$\pm$0.004 &    0.060$\pm$0.003 &    0.059$\pm$0.004 &    0.063$\pm$0.002 \\
$f_b$\ (OGLE)          &    0.096$\pm$0.004 &    0.096$\pm$0.003 &    0.097$\pm$0.003 &    0.092$\pm$0.002 \\
\hline
\label{tab:3l1s}
\end{tabular}
\end{center}
\end{table*}